\documentclass[epj-spec]{svjour}

\usepackage{bm}
\usepackage{amsmath}
\usepackage{amssymb}
\usepackage{graphicx}

\begin{document}

\title{Conductivity of disordered graphene at half filling}

\author{
  P.~M.~Ostrovsky\inst{1}\fnmsep\thanks{
    Also at L.~D.~Landau Institute for Theoretical Physics RAS,
    119334 Moscow, Russia.
  } \and
  I.~V.~Gornyi\inst{1}\fnmsep\thanks{
    Also at A.F.~Ioffe Physico-Technical Institute,
    194021 St.~Petersburg, Russia.
  } \and
  A.~D.~Mirlin\inst{1}\fnmsep\inst{2}\fnmsep\thanks{
    Also at Petersburg Nuclear Physics Institute,
    188300 St.~Petersburg, Russia.
  }
}

\institute{
 Institut f\"ur Nanotechnologie, Forschungszentrum Karlsruhe,
 76021 Karlsruhe, Germany \and
 Institut f\"ur Theorie der kondensierten Materie,
 Universit\"at Karlsruhe, 76128 Karlsruhe, Germany
}

\abstract{
We study electron transport properties of a monoatomic graphite layer
(graphene) with different types of disorder at half filling. We show that the
transport properties of the system depend strongly on the symmetry of disorder.
We find that the localization is ineffective if the randomness preserves one of
the chiral symmetries of the clean Hamiltonian or does not mix valleys. We
obtain the exact value of minimal conductivity $4e^2/\pi h$ in the case of
chiral disorder. For long-range disorder (decoupled valleys), we derive the
effective field theory. In the case of smooth random potential, it is a
symplectic-class sigma model including a topological term with $\theta = \pi$.
As a consequence, the system is at a quantum critical point with a universal
value of the conductivity of the order of $e^2/h$. When the effective time
reversal symmetry is broken, the symmetry class becomes unitary, and the
conductivity acquires the value characteristic for the quantum Hall transition.
}

\maketitle

\section{Introduction}

Almost 50 years ago, in a pathbreaking paper ``Absence of diffusion in certain
random lattices'' \cite{Anderson58} Anderson demonstrated that a quantum
particle may become localized by a random potential. In particular, in
non-interacting systems of two-dimensional (2D) geometry even weak disorder
localizes all electronic states \cite{Abrahams79,Gorkov79}, thus leading to the
exactly zero conductivity, $\sigma = 0$.

Recent breakthrough in graphene fabrication \cite{Novoselov04} and subsequent
transport experiments \cite{Novoselov05,Zhang,Novoselov06,Morozov06,Zhang06}
revealed remarkable electronic properties of this material
\cite{Wallace,AndoReview}. One of the most striking experimental observation
is the minimal conductivity $\sigma$ of order $e^2/h$ observed in undoped
samples and staying almost constant in a wide range of temperatures $T$ from
$300$K down to $\sim 1$K. This behavior should be contrasted to
well-established results on the conductivity of 2D systems where Anderson
localization drives the system into insulating state at low $T$
\cite{Abrahams79,Gorkov79,McCann,AleinerEfetov,Altland06}. An apparently
$T$-independent value of $\sigma \sim e^2/h$ suggests that the system is close
to a quantum critical point and calls for a theoretical explanation. The aim of
this paper is to analyze what one should expect for conductivity from the
theoretical point of view. We will see that, in view of the unconventional
character of the graphene spectrum, the nature of disorder is crucially
important.

In this paper we consider the conductivity of a single-layer graphene at half
filling, $\varepsilon = 0$. The Drude conductivity
\cite{ShonAndo,OurPRB} has the value $\sigma = 4e^2/\pi h$ at
this point. Since this value is of the order of conductance quantum, the
localization effects should, in principle, become strong at half filling. At
lowest temperatures $T$, the system with generic disorder preserving
time-reversal symmetry falls into the orthogonal symmetry class
\cite{McCann,AleinerEfetov,Altland06} implying strong localization regime
\cite{AleinerEfetov}. In this situation, the conductivity at half filling
should have a pronounced temperature dependence and get strongly suppressed
with lowering $T$. This conclusion remains valid for generic disorder also
if the time-reversal symmetry is broken (e.g. by magnetic impurities), so that
the symmetry is unitary \cite{Altland06}. Remarkably, this is not what is
observed in the experiment \cite{Novoselov05,Zhang}. According to what was said
above, this might only happen, if at all, for a particular type of disorder.

We identify two broad classes of randomness in graphene --- chiral disorder
(preserving chiral symmetry of clean graphene) and long-range disorder (not
mixing the two valleys) --- leading to the lack of localization and emergence
of quantum criticality and associated universal minimal conductivity
\cite{OurPRB,OurFuturePRL}.

\section{The Model}

The tight-binding Hamiltonian of clean graphene is a $4 \times 4$ matrix
operating in the AB space of two sublattices (Pauli matrices $\sigma_i$) and in
the $K$--$K'$ space of the valleys (Pauli matrices $\tau_i$). Therefore we
introduce the four-component wave function
\begin{equation}
 \Psi
  = \{\phi_{AK}, \phi_{BK}, \phi_{BK'}, \phi_{AK'}\}^T.
 \label{4D}
\end{equation}
In this representation the Hamiltonian has the form
\begin{equation}
 H
  = v_0 \tau_3 \bm{\sigma}\mathbf{k}.
 \label{ham}
\end{equation}
Here $\tau_3$ is the third Pauli matrix in the $K$--$K'$ space and $\bm{\sigma}
= \{\sigma_1, \sigma_2\}$ is the two-dimensional vector of Pauli matrices in
the AB space. The Fermi velocity in graphene is $v_0 \simeq 10^8$ cm/s. In
fact, the form of the Hamiltonian (\ref{ham}) is universal and does not rely on
the tight-binding approximation. The degeneracy of the spectrum in $K$ and $K'$
points is provided by the two-dimensional representation of the honeycomb
lattice symmetry group, while the expression (\ref{ham}) is the first-order
$k$-expansion near these points.

Let us analyze the symmetries of the clean graphene Hamiltonian (\ref{ham}).
First, the system is obviously uniform and isotropic. Any disorder considered
in this paper preserves these symmetries on average, so we do not pay much
attention to them here. Second, due to the two valley structure of the electron
spectrum the whole SU(2) symmetry group exists in an \emph{isospin} space of
the valleys. The generators of this group are \cite{McCann}
\begin{equation}
 \Lambda_x
  = \sigma_3 \tau_1,
 \qquad
 \Lambda_y
  = \sigma_3 \tau_2,
 \qquad
 \Lambda_z
  = \sigma_0 \tau_3.
 \label{Lambda-matrices}
\end{equation}
These three operators commute with the Hamiltonian and anticommute with each
other. Third, there are two more relevant symmetries of the clean Hamiltonian,
namely, time inversion operation (we denote it $T_0$) and chiral symmetry
($C_0$). Using various combinations of $T_0$, $C_0$, and rotations of the
isospin ($\Lambda_{0,x,y,z}$) one can easily construct a whole bunch of twelve
symmetry operations
\begin{align*}
 T_0: A
  &\mapsto \sigma_1 \tau_1 A^T \sigma_1 \tau_1, &
 C_0: A
  &\mapsto -\sigma_3 \tau_0 A \sigma_3 \tau_0, &
 CT_0: A
  &\mapsto -\sigma_2 \tau_1 A^T \sigma_2 \tau_1, \\
 T_x: A
  &\mapsto \sigma_2 \tau_0 A^T \sigma_2 \tau_0, &
 C_x: A
  &\mapsto -\sigma_0 \tau_1 A \sigma_0 \tau_1, &
 CT_x: A
  &\mapsto -\sigma_1 \tau_0 A^T \sigma_1 \tau_0, \\
 T_y: A
  &\mapsto \sigma_2 \tau_3 A^T \sigma_2 \tau_3, &
 C_y: A
  &\mapsto -\sigma_0 \tau_2 A \sigma_0 \tau_2, &
 CT_y: A
  &\mapsto -\sigma_1 \tau_3 A^T \sigma_1 \tau_3, \\
 T_z: A
  &\mapsto \sigma_1 \tau_2 A^T \sigma_1 \tau_2, &
 C_z: A
  &\mapsto -\sigma_3 \tau_3 A \sigma_3 \tau_3, &
 CT_z: A
  &\mapsto -\sigma_2 \tau_2 A^T \sigma_2 \tau_2.
\end{align*}
Note that the transposition changes sign of momentum operators.

Now we incorporate a disorder of the form $V_{ij} \sigma_i \tau_j$ in the
model, breaking some of the above symmetries. The average isotropy of the
disordered graphene implies that $\Lambda_x$ and $\Lambda_y$ symmetries of the
Hamiltonian are present or absent simultaneously. Below we combine them into a
single notation $\Lambda_\perp$, and proceed in the same way with $T_\perp$ and
$C_\perp$. In Table \ref{Tab:sym} we list all possible matrix structures of
disorder along with their symmetries.

\begin{table}

\caption{Symmetries of various disorders in graphene. The first five rows of
the table contain disorders preserving physical time inversion symmetry $T_0$.
The next four present structures violating $T_0$. Note that each row contains
``$+$'' either in the $\Lambda_z$ column or in one of the $C$ columns. This
means that each disorder structure from the first column --- when taken alone
--- produces a finite conductivity at the Dirac point.}
\label{Tab:sym}

\begin{center}
\begin{tabular}{c@{\qquad}c@{\,}cc@{\,}c@{\,}cc@{\,}c@{\,}cc@{\,}c@{\,}c}
\hline\hline
 Disorder structure &
 \makebox[0.7cm]{$\Lambda_\perp$} & \makebox[0.7cm]{$\Lambda_z$} &
 \makebox[0.7cm]{$T_0$} & \makebox[0.7cm]{$T_\perp$} & \makebox[0.7cm]{$T_z$} &
 \makebox[0.7cm]{$C_0$} & \makebox[0.7cm]{$C_\perp$} & \makebox[0.7cm]{$C_z$} &
 \makebox[0.7cm]{$CT_0$} & \makebox[0.7cm]{$CT_\perp$} & \makebox[0.7cm]{$CT_z$}
\\ \hline
 $\sigma_0 \tau_0$ &
 $+$ & $+$ &
 $+$ & $+$ & $+$ &
 $-$ & $-$ & $-$ &
 $-$ & $-$ & $-$
\\
 $\sigma_{\{1,2\}} \tau_{\{1,2\}}$ &
 $-$ & $-$ &
 $+$ & $-$ & $-$ &
 $+$ & $-$ & $-$ &
 $+$ & $-$ & $-$
\\
 $\sigma_{1,2} \tau_0$ &
 $-$ & $+$ &
 $+$ & $-$ & $+$ &
 $+$ & $-$ & $+$ &
 $+$ & $-$ & $+$
\\
 $\sigma_0 \tau_{1,2}$ &
 $-$ & $-$ &
 $+$ & $-$ & $-$ &
 $-$ & $-$ & $+$ &
 $-$ & $-$ & $+$
\\
 $\sigma_3 \tau_3$ &
 $-$ & $+$ &
 $+$ & $-$ & $+$ &
 $-$ & $+$ & $-$ &
 $-$ & $+$ & $-$
\\ \hline
 $\sigma_3 \tau_{1,2}$ &
 $-$ & $-$ &
 $-$ & $-$ & $+$ &
 $-$ & $-$ & $+$ &
 $+$ & $-$ & $-$
\\
 $\sigma_0 \tau_3$ &
 $-$ & $+$ &
 $-$ & $+$ & $-$ &
 $-$ & $+$ & $-$ &
 $+$ & $-$ & $+$
\\
 $\sigma_{1,2} \tau_3$ &
 $+$ & $+$ &
 $-$ & $-$ & $-$ &
 $+$ & $+$ & $+$ &
 $-$ & $-$ & $-$
\\
 $\sigma_3 \tau_0$ &
 $+$ & $+$ &
 $-$ & $-$ & $-$ &
 $-$ & $-$ & $-$ &
 $+$ & $+$ & $+$
\\ \hline\hline
\end{tabular}
\end{center}
\end{table}

Generally, the chiral symmetry implies that the Hamiltonian takes
block-off-diagonal form under a proper unitary transformation. For instance, a
generic disorder preserving $C_z$ symmetry can have only \emph{off-diagonal}
matrix elements in the AB space of sublattices \cite{Guruswamy}. Away from the
Dirac point ($\varepsilon \neq 0$), any chiral symmetry is broken by the
diagonal ($\sigma_0\tau_0$) energy term in the action. On the other hand, at
half filling the chiral symmetry plays crucial role in establishing the
transport properties of the system, see Sec.\ \ref{Sec:chiral}. If disorder
commutes with $\Lambda_z$ (long-range), the system splits into two copies
(decoupled valleys). The conductivity in this case is discussed in Sec.\
\ref{Sec:long-range}

In Table \ref{Tab:result} we list all possible situations when the symmetry
prevents the localization and leads to a finite conductivity at $\varepsilon=0$.
This happens if the system possesses chiral symmetry or the valleys are
decoupled, as we describe in detail below.

\begin{table}

\caption{Possible types of disorder in graphene leading to a finite minimal
conductivity. Each row of the table corresponds to a certain universality class
of the problem. In the first column we give an example of the disorder that
belongs to this class (RMF is an external random magnetic field). The relevant
symmetries are placed in the second column. The symmetry class itself is
indicated in the third column according to the classification of Altland and
Zirnbauer \cite{Zirnbauer,AltlandZirnbauer,AltlandSimonsZirnbauer}. The fourth
column displays the dominant non-trivial term in the corresponding sigma model.
This term is responsible for the absence of localization. There are three
possibilities \cite{FendleyLecNotes}: (i) Gade term \cite{Gade93,GadeWegner91},
(ii) Wess-Zumino-Witten term \cite{Witten,AltlandSimonsZirnbauer}, or (iii)
Pruisken $\theta$-term \cite{Pruisken} with $\theta = \pi$. The value of the
minimal conductivity is given in the last column. The first five rows contain
chiral-symmetric types of disorder [ripples ($\sigma_{1,2} \tau_0$) can be
added to any of them]. They all yield the minimal conductivity
$4e^2/\pi h$ (up to power-law corrections in a weak disorder strength for $C_z$
chirality, hence ``$\approx$''), see Sec.\ \protect\ref{Sec:chiral}. The last
four rows correspond to the case of decoupled valleys (long-range disorder),
see Sec.\ \protect\ref{Sec:long-range}; from top to  bottom: random Dirac
vector potential, scalar potential, mass, and any of their combinations.
The value of the minimal conductivity depends on a particular symmetry of the
long-range disorder.}
\label{Tab:result}

\begin{center}
\begin{tabular}{ccccc}
\hline\hline
Disorder &
Symmetries              & Class          & Sigma model &
Conductivity \\
\hline
Vacancies, strong potential impurities &
$C_z$, $T_0$            & BDI            & Gade &
$\approx 4e^2/\pi h$ \\
Vacancies + RMF &
$C_z$                   & AIII           & Gade &
$\approx 4e^2/\pi h$ \\
$\sigma_3\tau_{1,2}$ disorder &
$C_z$, $T_z$            & CII            & Gade &
$\approx 4e^2/\pi h$ \\
Dislocations &
$C_0$, $T_0$            & CI             & WZW &
$4e^2/\pi h$ \\
Dislocations + RMF &
$C_0$                   & AIII           & WZW &
$4e^2/\pi h$ \\
\hline
Ripples, RMF &
$\Lambda_z$, $C_0$      & $2 \times$AIII & WZW &
$4e^2/\pi h$ \\
\hline
Charged impurities &
$\Lambda_z$, $T_\perp$  & $2 \times$AII  & $\theta = \pi$ &
$4\sigma_{Sp}^{**}$ \\
random Dirac mass: $\sigma_3\tau_0$, $\sigma_3\tau_3$ &
$\Lambda_z$, $CT_\perp$ & $2 \times$D    & $\theta = \pi$ &
$4e^2/\pi h$ \\
Charged impurities + (RMF, ripples) &
$\Lambda_z$             & $2 \times$A    & $\theta = \pi$ &
$4\sigma_U^*$ \\
\hline\hline
\end{tabular}
\end{center}
\end{table}

\section{Chiral disorder}
\label{Sec:chiral}

The special class of disorder that we will consider in this section is the
randomness that preserves one of the chiral symmetries of the clean graphene
Hamiltonian \cite{OurPRB}. Some possible realizations of such type of disorder
are listed in Table \ref{Tab:result}. A peculiar behavior of 2D systems with
chiral disorder with respect to localization effects has been demonstrated by
Gade and Wegner \cite{GadeWegner91,Gade93}. They considered a random hopping
problem on a square lattice and showed that at zero energy, where the system
possesses the chiral symmetry, the RG $\beta$-function of the corresponding
sigma model vanishes to all orders in the inverse conductivity, implying that
the conductivity is not renormalized. This absence of usual infrared-singular
corrections to the conductivity due to Cooperon and diffuson loops can be
attributed to the fact that the ``antilocalizing'' interference corrections to
the density of states cancel the localization corrections to the diffusion
coefficient. The states at the band center $\varepsilon = 0$ are delocalized
and the conductivity $\sigma(\varepsilon = 0)$ takes a finite value depending
on the disorder strength. According to the classification of Refs.\
\cite{Zirnbauer,AltlandZirnbauer,AltlandSimonsZirnbauer}, the system studied in
Refs.\ \cite{GadeWegner91,Gade93} belongs to the chiral orthogonal symmetry
class BDI if the time-reversal symmetry is preserved and to the chiral unitary
symmetry class AIII otherwise. The peculiarity of the problem we are
considering is the Dirac dispersion of carriers. This will allow us to prove
below a statement that is still stronger than that of Gade and Wegner: we will
show that for certain types of chiral disorder all disorder-induced
contributions to conductivity cancel and $\sigma(\varepsilon = 0)$ takes the
universal value $4e^2/\pi h$.

\subsection{$C_0$ chirality}

Let us consider the disorder which preserves the $C_0$-chirality, $H =
-\sigma_3 H \sigma_3$. The random part of the Hamiltonian  contains matrices
$\sigma_{1,2}\tau_{3}$, $\sigma_{1,2}\tau_{1,2}$, and $\sigma_{1,2}\tau_0$.
According to Ref.\ \cite{BernardLeClair}, the random Dirac Hamiltonians
preserving the $C_0$ chirality and violating the $T_0$ symmetry belong to the
chiral symmetry class AIII, while the combination of $C_0$ chirality and $T_0$
symmetry drives the system into the Bogolyubov -- de Gennes symmetry class CI.
In both cases, the low-energy theory is affected
\cite{NersesyanTsvelikWenger,AltlandSimonsZirnbauer} by the presence of the
Wess-Zumino-Witten term \cite{Witten} in the action.

Let us turn to the calculation of the conductivity at half filling, assuming
the $C_0$ disorder. The conductivity is given by the Kubo formula
\begin{equation}
 \sigma
  = -\frac 1{2\pi\hbar}\mathop{\mathrm{Tr}} \Bigl[
      j^x ( G^R - G^A )
      j^x ( G^R - G^A )
    \Bigr].
 \label{KuboFull}
\end{equation}
Here `Tr' operation implies matrix trace and the spatial integration. Now we
use the identity
\begin{equation}
 \sigma_3 G^{R(A)}(\varepsilon) \sigma_3
  = - G^{A(R)}(-\varepsilon)
 \label{GRA}
\end{equation}
valid when disorder preserves $C_0$-chirality. This allows us to trade all
advanced Green functions in Eq.\ (\ref{KuboFull}) for retarded ones and thus to
present the conductivity at zero energy in terms of retarded Green functions.
Further, we exploit the following important relation between the components of
the current operator
\begin{equation}
 \sigma_3 j^x
  = -j^x \sigma_3
  = i j^y,
 \label{current-x-y}
\end{equation}
which is the consequence of the Dirac spectrum. At this point, our problem
differs from that considered by Gade and Wegner \cite{GadeWegner91,Gade93} who
dealt with a bipartite \emph{square} lattice with a non-linear electronic
spectrum.

The transformations Eqs.\ (\ref{GRA}) and (\ref{current-x-y}) allow us to cast
the Kubo formula in the following form:
\begin{equation}
 \sigma(\varepsilon=0)
  = -\frac 1{\pi\hbar} \sum_{\alpha = x,y}
    \mathop{\mathrm{Tr}} \Bigl[
      j^\alpha G^R
      j^\alpha G^R
    \Bigr].
 \label{KuboRR}
\end{equation}
At first glance, this expression is zero due to gauge invariance. Indeed, the
right-hand side of Eq.\ (\ref{KuboRR}) is proportional to the second derivative
of the partition function $Z[\mathbf{A}] = \mathop{\mathrm{Tr}} \log
G^R[\mathbf{A}]$ (or, equivalently, first derivative of the current
$\mathop{\mathrm{Tr}} j^\alpha G^R[\mathbf{A}]$) with respect to the constant
vector potential $\mathbf{A}$. The gauge invariance implies that a constant
vector potential does not affect gauge-invariant quantities like the partition
function or the current, so that the derivative is zero. This argument is,
however, not correct for the zero-order diagram (the one with no disorder
included), in view of the quantum anomaly \cite{Schwinger,Peskin}. The explicit
calculation then yields
\begin{equation}
 \sigma
  = -\frac{8 e^2 v_0^2}{\pi\hbar} \int \frac{d^2k}{(2\pi)^2}\,
    \frac{\delta^2}{(v_0^2 k^2 + \delta^2)^2}
  = \frac{4 e^2}{\pi h}.
 \label{universal}
\end{equation}
Here $\delta$ is an infinitesimal imaginary part in the denominator of the
Green function. Physically, it has a meaning of the electron lifetime or,
alternatively, a dephasing rate, and can be thought of as modelling processes
of escape of electrons in some reservoir or some dephasing mechanism. Models
with such a uniform constant value of $\delta$ were used in the literature to
imitate dephasing in quantum dots, see e.g. Ref.\ \cite{Efetov}. The
corrections to Eq.\ (\ref{universal}) are exponentially small in the disorder
strength. 

We note that the same universal value of the conductivity in the situation when
the only type of disorder is the Abelian random vector potential was previously
obtained in Ref.\ \cite{Ludwig}; for a certain type of the non-Abelian gauge
potential with Gaussian distribution, universal conductivity was also obtained
in Ref.\ \cite{Tsvelik}. The above derivation of the universal conductivity
remains valid for the case when a magnetic field of an arbitrary strength is
applied: the vector potential $\mathbf{A}$ couples to the current, i.e. to the
matrices $\tau_3 \sigma_{1,2}$, thus preserving the chiral symmetry. In this
context, it is worth mentioning the result of Hikami, Shirai, and Wegner
\cite{HikamiShiraiWegner93} who found that the longitudinal conductivity in the
center of the lowest Landau level of the chiral-disordered 2D electron gas is
equal exactly to $\sigma = 4e^2/\pi h$ in the limit of very strong magnetic
field, when the Landau level mixing can be neglected.

\subsection{$C_z$ chirality}

In this subsection we consider the disorder which preserves the $C_z$
chirality, $H = -\sigma_3 \tau_3 H \sigma_3 \tau_3$. The random part of the
Hamiltonian may then contain matrices $\sigma_3 \tau_{1,2}$, $\sigma_{1,2}
\tau_3$, $\sigma_{1,2} \tau_0$, and $\sigma_{0} \tau_{1,2}$. Random Dirac
Hamiltonians \cite{BernardLeClair} preserving the $C_z$ chirality and violating
the $T_0$ symmetry belong to the chiral unitary symmetry class AIII.  WZW
terms \cite{Witten} from the two valleys cancel out
\cite{AltlandSimonsZirnbauer} in this case. The combination of $C_z$ chirality
and the time reversal invariance $T_0$ corresponds to the chiral orthogonal
symmetry class BDI. Finally, the combination of $C_z$ chirality and $T_z$
symmetry falls into the chiral symplectic symmetry class CII. In all these
situations, the resulting theory is of the Gade type
\cite{GadeWegner91,Gade93}.

Let us turn to the conductivity at half filling for a generic disorder
preserving the $C_z$ chirality. The proof of the universality of the
conductivity based on gauge-invariance argument does not work now since the
$C_z$ chirality transformation of the Green's function generates the new vector
vertices $j^{x,y}\tau_3$ instead of physical currents. Nevertheless, for weak
disorder we find that the conductivity at half filling is still universal,
$\sigma(\varepsilon=0)=4e^2/\pi h$, up to corrections in powers of disorder
strength (see also \cite{Ryu}). To show this, we first calculate the
perturbative correction $\delta\sigma^{(1)}$ to the conductivity of a pure
system at the first order in disorder strength and find that it vanishes,
$\delta\sigma^{(1)}=0$. This implies that the conductivity at the Dirac point
does not depend on the ultraviolet cut-off of the theory and
can be presented as a series in the weak disorder strength (indeed the
second-order term is found to be finite) \cite{OurPRB}. Next, we recall that
for $C_z$ chirality, the RG $\beta$-function of the Gade-Wegner sigma model
\cite{GadeWegner91,Gade93} vanishes to all orders, so that there are no
singular quantum-interference corrections to $\sigma(\varepsilon=0)$ due to the
soft modes (impurity ladders). This proves that the expansion of
$\sigma(\varepsilon = 0)$ in powers of disorder converges. Thus for the case of
weak disorder the conductivity is universal with small corrections in powers of
the disorder strength (unlike in the case of the $C_0$ chirality, where the
corrections are nonperturbative in the disorder strength.)

Finally, let us mention the case of $C_\perp$ chirality. In an isotropic system
considered here, both $C_x$ and $C_y$ chiralities are expected to be present
simultaneously. This implies that the disordered Hamiltonian anticommutes with
both $\tau_1$ and $\tau_2$ and hence is proportional to $\tau_3$. Thus it is
split into two equivalent copies. This situation of decoupled valleys
corresponds to the case of long-range disorder and will be considered in Sec.\
\ref{Sec:long-range}.

\subsection{Conductivity at finite frequency or finite temperature}

In this subsection, we analyze the frequency dependence of the conductivity in
the case of chiral disorder. For completeness, we also keep a small level
width $\delta$ introduced above. It was crucial for the argument leading to
Eq.\ (\ref{universal}) that the system is exactly at half filling, $\varepsilon
= 0$. A non-zero frequency implies an integration over the energy range of the
width $\omega$, which breaks the chiral symmetry. When the frequency $\omega$
is much smaller than $\delta$, this effect is however negligible, the infrared
regularization is provided by $\delta$, and the universal result
(\ref{universal}) survives. In its turn, $\delta$ plays no role when
$\omega\gg\delta$: it is the frequency that serves as a dominant infrared
cutoff now. Then a very interesting new situation arises since $\omega$ plays a
twofold role, leading to two competing effects. On one hand, as discussed
above, the frequency drives the system away from the chiral-symmetric point and
thus restores localization. On the other hand, the frequency cuts off the
singular localization correction. Which of these effects wins? To answer this
question, one should compare $\omega$ with the level spacing in the
localization area, $\Delta_\xi(\varepsilon)$, where $\varepsilon\sim\omega$. It
turns out that $\Delta_\xi(\varepsilon)\sim \varepsilon\sim\omega$ for the
chiral disorder. This result is rather general and is only based on the fact
that the operator governing the flow of the system away from  criticality
couples to the energy in the action. Therefore the two competing effects of the
frequency (the localization and the infrared regularization) ``make a draw'' --
both of them are equally important. The system turns out to be, roughly
speaking, half way between the chiral fixed point and the conventional
symmetry. This results in a new universal (frequency-independent) value of the
conductivity $\sigma_\omega \sim e^2/h$ in the considered regime $\delta \ll
\omega$. More precisely, this value depends on the type of chirality and the
symmetry class of the system away from the Dirac point.

In the presence of interactions, the temperature $T$ also plays a twofold role,
similarly to the frequency. On one hand, it induces an averaging over the
energy window of the width $\sim T$, thus breaking the chiral symmetry and
``switching on'' the localization effects. On the other hand, the interaction
at finite $T$ generates a non-zero dephasing rate $\tau_{\phi}^{-1}(T)$ cutting
off the localization corrections. As we showed above, the level spacing
$\Delta_\xi(T)$ is $\sim T$, so that the result of the competition of these two
effects depend on the value of $T\tau_\phi(T)$. It is natural to assume that
$\tau_\phi^{-1}(T) \sim T$ \cite{AleinerEfetov,OurPRB},
as follows from the result of Ref.\ \cite{AAK} for a
two-dimensional diffusive metal with Coulomb interaction at $\sigma \sim
e^2/h$. Then the behavior of the conductivity at low $T$ will be qualitatively
analogous to that for the case of finite frequency.

More realistically, one can think about a situation when the
disorder is predominantly chiral (say, due to ripples and dislocations),
but the chiral symmetry is
slightly broken, e.g., by weak short-range potential disorder.
Then the above consideration will be applicable in the
parametrically broad range of $T$; at the lowest temperatures,
the chirality-breaking effects will drive the system into the
strong localization regime.

\section{Long-range disorder}
\label{Sec:long-range}

In the previous section we have shown that, if one of the chiral symmetries of
clean graphene is preserved by disorder, the conductivity at half filling is
not affected by localization and is equal to $4 e^2/\pi h$. While various types
of randomness in graphene (in particular, dislocations, ripples, or strong
point-like defects) do belong to the chiral type, the experimentally observed
value of $\sigma$ is larger by a factor $\sim 3$, suggesting a different type
of criticality. In this section we consider another broad class of randomness
in graphene --- long-range disorder \cite{OurFuturePRL} --- commuting with
$\Lambda_z$ (see Table \ref{Tab:sym}). We will show the possibility of a new
critical state when valleys are decoupled. This case has a particular
experimental relevance if the conductivity is dominated by charged impurities.
The ripples \cite{Morpurgo06,Meyer07} belong to this type of randomness as
well. Numerical simulations of graphene with long-range random potential
\cite{Nomura06,Beenakker_numerics} provide an evidence in favor of a
scale-invariant conductivity.

The characteristic feature of the long-range disorder is the absence of valley
mixing due to the lack of scattering with large momentum transfer. This allows
us to describe the system in terms of a single-valley Dirac Hamiltonian with
disorder \cite{Ludwig,NersesyanTsvelikWenger,Bocquet},
\begin{equation}
 H
  = v_0 \bm{\sigma}\mathbf{k} + \sigma_\mu V_\mu(\mathbf{r}).
 \label{ham1}
\end{equation}
Here disorder includes random scalar ($V_0$) and vector ($V_{1,2}$) potentials
and random mass ($V_3$). The Hamiltonian (\ref{ham1}) was considered in Ref.
\cite{Ludwig} as a model for quantum Hall transition.

The clean single-valley Hamiltonian (\ref{ham1}) obeys \cite{Ludwig,SuzuuraAndo} the
effective time-reversal invariance $H = \sigma_2 H^T\sigma_2$. This symmetry
($T_\perp$) is not the physical time-reversal symmetry ($T_0$): the latter
interchanges the nodes and is of no significance in the absence of inter-node
scattering. Below we refer to the effective time-reversal symmetry in a single
node as the TR symmetry. If the only disorder is random scalar potential, the
TR invariance is not broken and the system falls into the symplectic symmetry
class AII, see Refs.\ \cite{Ludwig,AleinerEfetov,Altland06,McCann}. The standard
realization of such symmetry is a system with spin-orbit coupling; in the
present context the role of spin is played by the sublattice space.

\subsection{Unitary class}

We start with a more generic case of the unitary symmetry (class A). The TR
invariance is broken as soon as a (either random or non-random) mass or vector
potential is included, in addition to the scalar potential. To analyze the
critical behavior of the system, we derive the nonlinear sigma model for the
single-node Dirac fermions \cite{OurFuturePRL} at finite energy $\varepsilon$.
This model is formulated in terms of the $4 \times 4$ supermatrix field $Q$
operating in Bose-Fermi (BF) superspace and bearing the retarded-advanced (RA)
structure \cite{Efetov}. The sigma-model action reads \cite{OurFuturePRL}
\begin{equation}
 S[Q]
  = \frac{1}{4} \mathop{\mathrm{Str}} \left[
       -\frac{\sigma_{xx}}{2} (\nabla Q)^2
       +\biggl( \sigma_{xy} + \frac{1}{2} \biggr)
         Q \nabla_x Q \nabla_y Q
    \right]
  \equiv S_1[Q] + i S_2[Q].
\label{final-action}
\end{equation}
Here `Str' operation includes the matrix supertrace and the spatial
integration. The parameters $\sigma_{xx}$ and $\sigma_{xy}$ of the action
are the longitudinal and Hall conductivities (in units $e^2/h$) given
by the corresponding Kubo expressions for a single node. In the presence of
the constant mass $m$, they acquire the values
\begin{gather}
 \sigma_{xx}
  = \frac{1}{2\pi} \left[
      1 + \frac{\varepsilon^2 + \gamma^2 - m^2}{2\gamma} f(\varepsilon, m)
    \right],
\qquad
 \sigma_{xy}
  = -\frac{m}{2\pi} \bigl[ f(\varepsilon,m) + f(m,\varepsilon) \bigr], \\
 f(x, y)
  = \frac{1}{x} \left[
      \arctan \frac{x + y}{\gamma} + \arctan \frac{x - y}{\gamma}
    \right],
\end{gather}
where $\gamma$ is the elastic scattering rate.
Note that if the global $T_0$ symmetry involving both the valleys
is preserved, the observable
Hall conductivity is absent due to the cancellation between
the contributions of the two valleys.

The imaginary part of the action, $i S_2[Q]$, is determined by the
well-known topological invariant $\mathop{\mathrm{Str}} ( Q
\nabla_x Q \nabla_y Q) \equiv 8 i \pi N[Q]$ on the sigma-model
manifold \cite{Pruisken}. Possible values of this functional are
integer multiples of $8 \pi i$. In Eq.\ (\ref{final-action}), the
topological term is equal to $i\theta N[Q]$, with the angle
$\theta = 2 \pi \sigma_{xy} + \pi$. In graphene the mass $m$ is
absent, so that $\sigma_{xy}=0$. Thus, the topological angle is
$\theta = \pi$. The theory (\ref{final-action}) is then exactly on
the critical line of the quantum Hall transition
\cite{Pruisken,Khmelnitskii}, in agreement with the arguments of
Ref.\ \cite{Ludwig}. We thus conclude that graphene with a generic
(TR-breaking) long-range disorder is driven into the quantum Hall
critical point (i.e. we have a ``quantum Hall effect without Landau levels"
 \cite{Haldane}), with the conductivity $4\sigma^*_U$. The factor 4
here accounts for the spin and valley degeneracy. The value
$\sigma^*_U$ is known to be in the range $\sigma^*_U \simeq 0.5 -
0.6$ $e^2/h$ from numerical simulations
\cite{Quantum-Hall_critical}. A schematic scaling function in this
case is shown in Fig.\ \ref{Fig:beta}a. While formally this
conclusion holds for any energy $\varepsilon$, in reality it only
works near half filling; for other $\varepsilon$ the quantum Hall
critical point would only be reached for unrealistic temperatures
and system sizes. It is worth noting that the sigma model Eq.\
(\ref{final-action}) is rigorously derived only for $\sigma_{xx}
\gg 1$, whereas $\sigma_{xx} \sim 1$ at $\varepsilon = 0$. This
should not, however, affect the universal critical behavior of the
theory governed by symmetry and topology of the problem.
An alternative route from the Dirac anomaly to a topological term employs
non-Abelian bosonization, see Refs. \ \cite{AltlandSimonsZirnbauer,Bocquet}.

If a uniform transverse magnetic field is applied, the topological angle
$\theta$ becomes energy-dependent. However, at zero energy, where $\sigma_{xy}
= 0$, its value remains unchanged, $\theta = \pi$. This implies the emergence
of the half-integer quantum Hall effect, with a plateau transition point at
$\varepsilon = 0$. An alternative explanation \cite{Novoselov05}
of the half-integer quantum Hall
effect is based on the concept of the Berry phase $\pi$ in graphene
\cite{SuzuuraAndo}. This relates the Berry phase and the anomalous topological
angle $\theta = \pi$.

In the above, we have assumed a generic TR-breaking long-range disorder. For
specific kinds of disorder, additional $C$ or $CT$ symmetries may emerge at
zero energy. In particular, if the only disorder is a random vector potential
(ripples), the system belongs to the chiral class AIII, with $\sigma = 4e^2/\pi
h$, see Sec.\ \ref{Sec:chiral}. If only the random mass is present, the system
falls into class D at $\varepsilon = 0$. In this case the weak disorder is
known to be irrelevant \cite{Ludwig,NersesyanTsvelikWenger,Bocquet}, implying
that the conductivity is again $4e^2/\pi h$.

\subsection{Symplectic class}

Let us now turn to the case of preserved TR invariance, describing in
particular charged impurities. The system belongs then to the symplectic
symmetry class AII. The derivation of the $\sigma$-model starts with the
doubling of $\psi$ variables accounting for the TR symmetry \cite{Efetov}. Then
$Q$ is a $8\times 8$ matrix obeying an additional constraint of charge
conjugation $Q = \bar Q$.

Since the partition function of the symplectic model is real, the imaginary
part of the action $S_2$ can take one of the two possible values, $0$ or $\pi$.
The discreteness of $S_2$ suggests that it again should be proportional to a
topological invariant on the sigma-model manifold. A non-trivial topology may
arise only in the compact (fermion) sector of $Q$. The corresponding target
space is $\mathcal{M}_F = O(4n) / O(2n) \times O(2n)$, where $n$ is the number
of fermion species. While for the conventional (sufficient for the analysis of
conductivity) sigma model $n = 1$, larger values will arise if one considers
higher-order products of Green functions.
The topological invariant takes values from the homotopy group
\cite{ViroFuchs}
\begin{equation}
 \pi_2 \bigl[ \left. \mathcal{M}_F\right|_{n=1} \bigr]
  = \mathbb{Z} \times \mathbb{Z}, \qquad\qquad
 \pi_2 \bigl[ \left. \mathcal{M}_F\right|_{n\geq2} \bigr]
  = \mathbb{Z}_2.
 \label{pi2}
\end{equation}
The homotopy group in the case $n = 1$ is richer than for $n \geq 2$.
Nevertheless, $S_2$ may take only two non-equivalent values. Hence only a
certain $\mathbb{Z}_2$ subgroup of the whole $\mathbb{Z} \times \mathbb{Z}$
comes into play as expected: the phase diagram of the theory should not depend
on $n$. Possibility of the $\mathbb{Z}_2$ topological term in the 2D symplectic
sigma model was emphasized by Fendley in Refs.\ \cite{FendleyLecNotes,Fendley}.

\begin{figure}
\centerline{
\includegraphics[width=6cm]{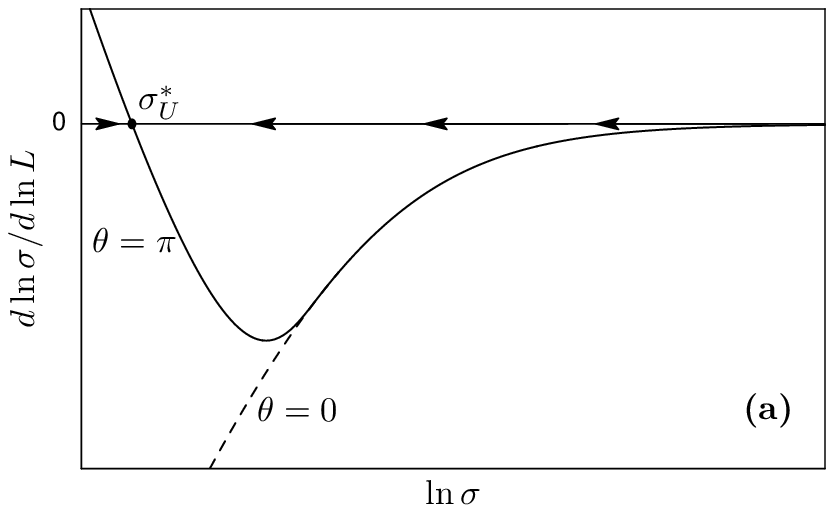}
\qquad
\includegraphics[width=6cm]{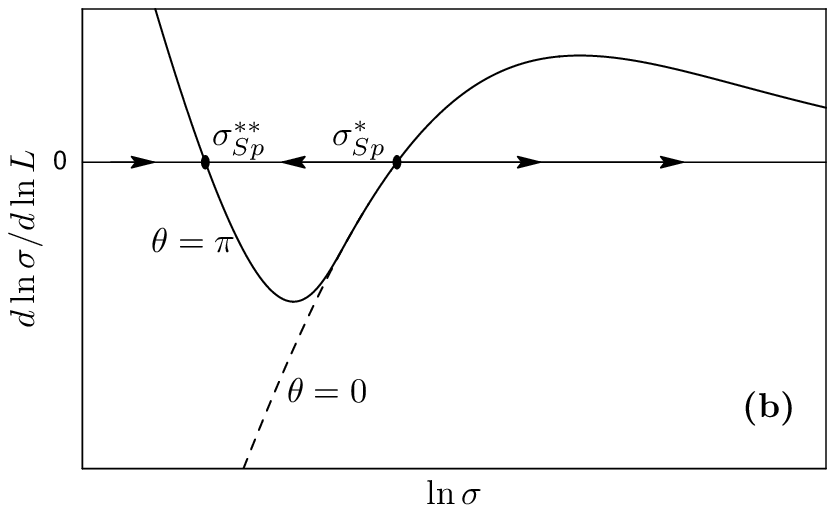}
}
\caption{Schematic scaling functions for (a) unitary and (b) symplectic
universality class with topological term $\theta = \pi$. The ordinary case
$\theta = 0$ is shown by dashed lines.}
\label{Fig:beta}
\vspace*{-0.5cm}
\end{figure}

To demonstrate the emerging topology explicitly and to calculate the
topological invariant, let us analyze the case $n = 1$ in more detail. The
generators of the compact sector are Hermitian skew-symmetric $4 \times 4$
matrices anticommuting with $\Lambda \equiv \rho_3$: $\rho_1 \mu_2$, $\rho_2
\mu_0$, $\rho_2 \mu_1$, and $\rho_2 \mu_3$. Here  $\rho_i$ and $\mu_i$ are
Pauli matrices in RA and TR space respectively. These generators split into two
mutually commuting pairs, each generating a 2-sphere (``diffuson'' and
``Cooperon'' sphere). Simultaneous inversion of both spheres leaves $Q$ intact.
Hence the compact sector of the model is the manifold $(\mathcal{S}_2 \times
\mathcal{S}_2) / \mathbb{Z}_2$. Thus two topological invariants, $N_{1,2}[Q]$,
counting the
covering of each sphere, emerge in accordance with Eq.\ (\ref{pi2}). The most
general topological term is $\theta_1 N_1 + \theta_2 N_2$. Due to the TR
symmetry, the action is invariant under interchanging the diffuson and Cooperon
spheres, which yields $\theta_1 = \theta_2 \equiv \theta$ where $\theta$ is
either $0$ or $\pi$. The explicit calculation of the $n=1$ topological term
yields
\begin{equation}
 i S_2[Q]
  = \frac{\epsilon_{\alpha\beta}}{8} \mathop{\mathrm{Str}} \bigl[
      (\Lambda \pm 1) \mu_2 u_\alpha u_\beta
    \bigr],
 \label{theta-term}
\end{equation}
where $\mathbf{u} = T \nabla T^{-1}$ and $Q=T^{-1} \Lambda T$. The sign
ambiguity here does not affect any observables. Equation (\ref{theta-term}) can be
cast in the form $i S_2[Q] = i \pi (N_1[Q] + N_2[Q])$, thus yielding  $\theta =
\pi$. If the TR symmetry is broken, the Cooperon modes are frozen and the
manifold is reduced to a single diffuson sphere ($Q$, and hence $\mathbf{u}$,
commutes with $\mu_2$) with $i S_2 = i \pi N_1[Q]$. In this case, $N_1[Q] =
(\epsilon_{\alpha\beta}/8 i \pi) \mathop{\mathrm{Str}}(\Lambda \mu_2 u_\alpha
u_\beta) = (i/16 \pi) \mathop{\mathrm{Str}} (Q \nabla_x Q \nabla_y Q \mu_2)$.
In the case $n>1$, the topological term is $i S_2[Q] = i \pi N[Q]$ with $N[Q]
\in \mathbb{Z}_2 \equiv \{0,1\}$, cf.\ Eq.\ (\ref{pi2}). 
The statement $\theta=\pi$ follows from the fact that $Q$ configurations 
of the $n=1$ theory which have an odd topological charge $N[Q]$ and yield a
negative partition function, are inherited by the $n>1$ theory by virtue of
natural embedding of the corresponding manifolds. The explicit form Eq.\
(\ref{theta-term}) is, however, not applicable in the $n>1$ case, since the
manifold no longer decomposes into two spheres.

An ordinary symplectic theory with no topological term exhibits metal-insulator
transition  at $\sigma^*_{Sp} \approx 1.4$ $e^2/h$ \cite{Schweitzer}. If the
conductivity is smaller than this critical value, the localization drives the
system into insulating state, while in the metallic phase, $\sigma >
\sigma^*_{Sp}$, antilocalization occurs. Using the analogy with the quantum
Hall transition in the unitary class \cite{Pruisken,Khmelnitskii}, we argue
that the topological term with
$\theta=\pi$ suppresses localization effects when the conductivity is small,
leading to appearance of a new attracting fixed point at $\sigma^{**}_{Sp}$.
The position of the metal-insulator transition, $\sigma^*_{Sp}$, is also
affected by the topological term. However, we believe that its change is
negligible: the instanton correction to the scaling function at large
conductivity is exponentially small \cite{Pruisken}, and the value of the
exponential factor $e^{-2 \pi \sigma}$ is still extremely small at $\sigma =
\sigma^*_{Sp}$. A plausible scaling of the conductivity in the symplectic case
with $\theta = \pi$ is sketched in Fig.\ \ref{Fig:beta}b. The existence and
position of the new critical point can be verified numerically. Recent
simulations of graphene \cite{Nomura06,Beenakker_numerics} indeed demonstrate
the stability of the conductivity in the presence of long-range disorder. Of
course, in reality there will be always a weak inter-valley scattering, which
will establish the localization at lowest $T$, in agreement with
\cite{AleinerEfetov,Altland06}. However, the approximate quantum criticality
will hold in a parametrically broad range of $T$.

Before concluding, let us briefly discuss this crossover to localization in
graphene with charged impurities. The dominant diagonal ($\sigma_0 \tau_0$)
scattering amplitude does not participate in the ultra-violet renormalization
\cite{Ludwig,NersesyanTsvelikWenger,AleinerEfetov,OurPRB},
in contrast to the case of a finite-range intra-valley disorder. The
reason is the long-range type of correlations ($\propto q^{-2}$ in momentum
space) that resulted in the absence of ultra-violet corrections to the linear
conductivity in Refs.\ \cite{Ando06,Nomura06,OurPRB,Khveshchenko}. 
Near half filling, the 
scattering rate is $\gamma \sim h v_0 n_\text{imp}^{1/2}$, where $n_\text{imp}$
is the concentration of Coulomb impurities and the interaction parameter $r_s$
is set to unity. The inter-valley processes occur
with a much smaller rate $\gamma' \sim a^2 n_\text{imp} \gamma$,
where $a$ is the lattice constant. Note that the value
of $\gamma'$ is only weakly corrected due to ultra-violet renormalization, unlike
in the model of white-noise disorder \cite{AleinerEfetov}. Assuming as above
$\tau_\phi^{-1} \sim T$, the crossover to strong localization (orthogonal
symmetry class AI) takes place at temperatures of the order of $\gamma'$.
Using realistic parameters from Ref.\ \cite{Novoselov05}, we find that the
symplectic quantum criticality characterized by a scale-invariant conductivity
$\sigma^{**}_{Sp}$ holds down to $10-50$ mK. Therefore, the model of long-range
charged impurities is capable of describing simultaneously the two most
striking experimental observations \cite{Novoselov05,Zhang}:
the linear density dependence of the conductivity \cite{Ando06,Nomura06}
and the existence of a universal minimal conductivity at the Dirac point
\cite{OurFuturePRL} in a broad temperature range.

\section{Conclusions}

To summarize, we have studied electron transport properties of a disordered
graphene layer. We have shown that the nature of disorder is of crucial
importance for the behavior of the conductivity. At half filling, the
conductivity is of the order of $e^2/h$ if the randomness preserves one of the
chiral symmetries of the clean Hamiltonian or does not mix the valleys.

For the case of chiral disorder, the exact value of the conductivity is still
determined by the nature of the infrared cutoff, which may depend on the physical
setup. We have analyzed in detail the situation when this cutoff is provided by
the level width $\delta$ or by the frequency $\omega$. In the former case the
conductivity takes a universal value $4 e^2/\pi h$, while in the latter case it
shows a more complex behavior still being of the order of conductance quantum.

In the case of the long-range (no inter-valley scattering) disorder, graphene
also shows quantum criticality at half filling. If the effective TR symmetry of the
single-valley system is preserved (e.g. when Coulomb scatterers are the
dominant disorder), the relevant theory is the symplectic sigma model with
topological angle $\theta=\pi$ and the minimal conductivity takes the
universal value $4\sigma^{**}_{Sp}$. If the TR symmetry is broken (e.g. by
effective random magnetic field due to ripples or by external magnetic field),
the system falls into the universality class of the quantum Hall critical
point, with another universal value $4\sigma^*_U$.

\section{Acknowledgments}
We thank V.~Cheianov, D.I.~Diakonov, F.~Evers, A.~Geim, I.A.~Gruzberg, H.~von
L\"ohneysen, A.W.W.~Ludwig, Y.~Makhlin, S.V. Morozov, A.F. Morpurgo, C. Mudry,
K.S. Novoselov, S.~Ryu, V. Serganova, M.A.~Skvortsov, A.G.~Yashenkin, and
I.~Zakharevich for valuable discussions and comments. The work was supported by
the Center for Functional Nanostructures and the Schwerpunktprogramm
``Quanten-Hall-Systeme'' of the Deutsche Forschungsgemeinschaft. The work of
PMO was supported by the Russian Foundation for Basic Research under grant No.\
04-02-16348 and by the Russian Academy of Sciences under the program ``Quantum
Macrophysics.'' ADM acknowledges hospitality of the
Kavli Institute for Theoretical Physics at Santa Barbara and partial support by
the National Science Foundation under Grant No.\ PHY99-07949.
The work of IVG, conducted as a part of the project ``Quantum
Transport in Nanostructures'' made under the EUROHORCS/ESF EURYI Awards scheme,
was supported by funds from the Participating Organizations of EURYI and the EC
Sixth Framework Programme. IVG is grateful to the organizers of the workshop
COQUSY06 for a friendly and stimulating atmosphere during the Graphene
Conference and acknowledges the hospitality of the Max Planck Institute for the
Physics of Complex Systems in Dresden.
IVG is also indebted to the museum guide of the COQUSY06-excursion
at the Green Vault in Dresden for his layman's remark: ``Graphene?
I know: this is like the quantum Hall effect!''

\end{document}